\documentclass[aip,graphicx]{revtex4-1}

\draft 

\begin{document}

\title{Dynamical symmetry breaking in geometrodynamics} 

\author{Alcides Garat}
\affiliation{1. Instituto de F\'{\i}sica, Facultad de Ciencias,
Igu\'a 4225, esq. Mataojo, Montevideo, Uruguay.}

\date{December 18th, 2011}

\begin{abstract}
We will analyze through a first order perturbative formulation the local loss of symmetry when a source of electromagnetic and gravitational field interacts with an agent that perturbs the original geometry associated to the source. As the local gauge symmetry in Abelian or even non-Abelian field structures in four-dimensional Lorentzian spacetimes is displayed through the existence of local planes of symmetry that we will refer to as blades one and two, the loss of symmetry will be manifested by the tilting of these planes under the influence of an external agent. In this strict sense the original local symmetry will be lost. We will be able to prove in this way that the new blades at the same point will correspond ``after the tilting generated by perturbation'' to a new symmetry. The purpose of this paper is to show that the geometrical manifestation of local gauge symmetries is dynamic. Despite the fact that the local original symmetries will be lost, new symmetries will arise. A dynamic evolution of local symmetries will be evidenced. This result will produce a new theorem on dynamic symmetry evolution.
\end{abstract}


\maketitle 

\section{Introduction}

Over the past decades dynamic symmetry breaking has received much attention from numerous authors \cite{NJL}$^{-}$\cite{SW}. These studies focused mainly on the mechanisms of mass generation, where mass is considered as a result of different forms of breaking symmetries. In these works Quantum Field Theoretical techniques were used. In our work we will address the subject of dynamical symmetry breaking but from a geometrical and classical point of view. We must stress from the beginning that our goal is not mass generation but the generation of a change in curvature that will be responsible for the symmetry breaking. We assume the existence of classical sources of gravitational fields. The sources also generate electromagnetic fields and we limit ourselves to the Abelian case even though there is a direct extension to the non-Abelian situation. Later, we will focus on analyzing the relationship between our study and the works cited above. For this purpose we have to review the results found in previous works such as one the manuscript stated in reference \cite{A}. It was found that at every point in a curved four-dimensional Lorentzian spacetime where a non-null electromagnetic field is present, a tetrad can be built so that these vectors covariantly diagonalize the stress-energy tensor at every point in spacetime. These tetrads define two local orthogonal planes. Blade one is generated by a timelike and a spacelike vector. Blade two by the other two spacelike vectors. All the vectors on the local planes one are eigenvectors of the Einstein-Maxwell stress-energy tensor with the same eigenvalue. All the vectors on the local plane two are also eigenvectors of the stress-energy tensor with minus the eigenvalue of local plane one. These new tetrad vectors have in their construction two elements. One of their building blocks is the skeleton. Skeletons are built by using the extremal fields, and extremal fields are found by means of local duality rotations of electromagnetic fields in the Einstein-Maxwell case, for instance. Extremal fields are local electromagnetic gauge invariants in the Abelian electromagnetic case and therefore the skeletons are gauge invariants. The other tetrad construction element is the gauge vector. The gauge vectors are gauge per se, and contain gauge fields in their construction. We can then observe that these tetrad vectors are gauge dependent, locally. A local electromagnetic gauge transformation might be thought of alternatively as a different choice for the gauge vectors, since the gauge vectors are gauge per se, they are a choice, as long as the tetrad vectors do not become trivial. When these new choices are considered and we analyze the change in the unit normalized tetrad vectors under a local gauge transformation, we see that these vectors transform inside the local planes that they originally defined without leaving them. Keeping in the process the metric tensor invariant. It has been proved through detailed analysis \cite{A} case by case that the local groups of Abelian electromagnetic gauge transformations are isomorphic to the following local groups of tetrad transformations. On plane one the local group of electromagnetic gauge transformations is isomorphic to the group $SO(1,1)$ of tetrad boosts, plus two discrete transformations. One of the discrete transformations is the full inversion or just minus the identity in a two by two matrix. The other discrete transformation is called the ``switch'' and we can represent it by a two by two matrix with ones off diagonal and zeroes in the diagonal. This last tetrad transformation is not Lorentzian. These set of tetrad transformations make up a new group LB1 or Lorentz blade one group. On plane two the local group of electromagnetic gauge transformations is isomorphic to $SO(2)$ tetrad spatial rotations which for this specific purpose we called LB2 or Lorentz blade two group. By transitivity LB1 is isomorphic to LB2. This is a new result in group theory. Through the work carried out in this previous paper \cite{A} we found that locally the electromagnetic gauge group of transformations was isomorphic to tetrad Lorentz transformations in both orthogonal planes or blades, one and two \cite{SCH}. That is to say, isomorphic to local Lorentz transformations on both planes, independently. We keep in mind that the discrete switch on blade one is not a Lorentz transformation. Therefore, the symmetry represented by local electromagnetic gauge transformations can be thought of as Lorentz transformations of the tetrad unit vectors inside these blades. It is within the context provided by all the results found \cite{A} that we will argue the following statements. First, mass needs to be associated to a dynamic process of symmetry breaking or another process like the Higgs \cite{PH} mechanism when addressed from the point of view of the standard model where gravitational fields are not present. Second, the very notion of symmetry breaking in the context where symmetries are treated as conserved properties that might be broken with the ensuing mass generation, phenomenon that leads to the results enumerated in the previous list of standard model approaches, is reformulated in this manuscript. In our geometrical context where local gauge transformations are reinterpreted as local Lorentz tetrad transformations, symmetries are not treated as conserved properties themselves in the following sense. They are broken by the action of external geometrical agents meaning that the local planes of symmetry will be tilting as the evolution of the interaction takes place. The local orthogonal planes of diagonalization of the stress-energy tensor will be perturbed as the fields are perturbed by the external agent in a continuous way. They will be broken in such a way that there will be new planes or blades at every point so that new symmetries will arise as time evolves. These will be symmetries of analogous nature but on new local planes. Symmetries then will not be broken under this new point of view, they will evolve dynamically. Mass within the context of the standard model is generated through dynamical symmetry breaking or the Higgs mechanism, whereas in our context what is generated is a change in curvature. In section \ref{reissnordexample} we will introduce the subject of local symmetry in terms of Lorentz tetrad transformations on blades one and two taking the Reissner-Norstr\"{o}m geometry as an example. In section \ref{fopert} we will introduce the first order perturbative treatment of geometrodynamics and its relation to dynamic symmetry breaking. We will state a new theorem on the evolution of symmetries. Throughout our work we use the conventions used in paper \cite{MW}. In particular we use a metric with sign conventions -+++. The only difference in notation with \cite{MW} will be that we will call our geometrized electromagnetic potential $A^{\alpha}$, where $f_{\mu\nu}=A_{\nu ;\mu} - A_{\mu ;\nu}$ is the geometrized electromagnetic field $f_{\mu\nu}= (G^{1/2} / c^2) \: F_{\mu\nu}$.


\section{The Reissner-Nordstr\"{o}m geometry as an example}
\label{reissnordexample}

In this section we will study the Reissner-Norstr\"{o}m geometry example. The line element for this spacetime is given by the following expression,

\begin{eqnarray}
ds^{2} = - (1 - {2m \over r} + {q^{2} \over r^{2}})\: dt^{2} + (1 - {2m \over r} + {q^{2} \over r^{2}})^{-1}\: dr^{2} + r^{2}\:(d\theta^{2} + \sin^{2}\theta\:d\phi^{2})\ . \label{reissnord}
\end{eqnarray}

In this section we will introduce for this solution to the Einstein-Maxwell equations the conserved currents and conserved charges. We will also introduce the vectors that diagonalize locally and covariantly the stress-energy tensor and generate blades one and two. This example will enable us to introduce perturbations in a general case with more clarity in section \ref{fopert}. Perturbations which will be considered under the scope of a tetrad geometric interpretation of local gauge transformations. Tetrads become tools of primary importance, as local gauge symmetries are associated to structures that can be expressed in terms of these new tetrad vectors. We present first, the four tetrad vectors introduced in paper \cite{A} that locally and covariantly diagonalize the electromagnetic stress-energy tensor and define at every point in spacetime blades one and two.

\begin{eqnarray}
V_{(1)}^{\alpha} &=& \xi^{\alpha\lambda}\:\xi_{\rho\lambda}\:X^{\rho}
\label{V1}\\
V_{(2)}^{\alpha} &=&  \sqrt{-Q/2} \:\: \xi^{\alpha\lambda} \: X_{\lambda}
\label{V2}\\
V_{(3)}^{\alpha} &=& \sqrt{-Q/2} \:\: \ast \xi^{\alpha\lambda} \: Y_{\lambda}
\label{V3}\\
V_{(4)}^{\alpha} &=& \ast \xi^{\alpha\lambda}\: \ast \xi_{\rho\lambda}
\:Y^{\rho}\ .\label{V4}
\end{eqnarray}

As a reminder we briefly state that the original expression for the electromagnetic stress-energy tensor $T_{\mu\nu}= f_{\mu\lambda}\:\:f_{\nu}^{\:\:\:\lambda}
+ \ast f_{\mu\lambda}\:\ast f_{\nu}^{\:\:\:\lambda}$ is given in terms of the electromagnetic tensor $f_{\mu\nu}$ and its dual $\ast f_{\mu\nu}={1 \over 2}\:\epsilon_{\mu\nu\sigma\tau}\:f^{\sigma\tau}$. After a local duality transformation,

\begin{equation}
f_{\mu\nu} = \xi_{\mu\nu} \: \cos\alpha +
\ast\xi_{\mu\nu} \: \sin\alpha\ ,\label{dr}
\end{equation}

where the local scalar $\alpha$ is the complexion, we are able to establish the stress-energy in terms of the extremal field $\xi_{\mu\nu}$ and its dual. We can express the extremal field as,

\begin{equation}
\xi_{\mu\nu} = e^{-\ast \alpha} f_{\mu\nu}\ = \cos\alpha\:f_{\mu\nu} - \sin\alpha\:\ast f_{\mu\nu}.\label{dref}
\end{equation}

Extremal fields are essentially electric fields and they satisfy,

\begin{equation}
\xi_{\mu\nu} \ast \xi^{\mu\nu}= 0\ . \label{i0}
\end{equation}

Equation (\ref{i0}) is a condition imposed on (\ref{dref}) and then the explicit expression for the complexion emerges $\tan(2\alpha) = - f_{\mu\nu}\:\ast f^{\mu\nu} / f_{\lambda\rho}\:f^{\lambda\rho}$. Like antisymmetric fields in a four-dimensional Lorentzian spacetime, the extremal fields also verify the identity,

\begin{eqnarray}
\xi_{\mu\alpha}\:\xi^{\nu\alpha} -
\ast \xi_{\mu\alpha}\: \ast \xi^{\nu\alpha} &=& \frac{1}{2}
\: \delta_{\mu}^{\:\:\:\nu}\ Q \ ,\label{i1}
\end{eqnarray}

where $Q=\xi_{\mu\nu}\:\xi^{\mu\nu}=-\sqrt{T_{\mu\nu}T^{\mu\nu}}$
according to equations (39) in \cite{MW}. $Q$ is assumed not to be zero,
because we are dealing with non-null electromagnetic fields. Condition (\ref{i0}) can be proved and through the use of the general identity,

\begin{eqnarray}
A_{\mu\alpha}\:B^{\nu\alpha} -
\ast B_{\mu\alpha}\: \ast A^{\nu\alpha} &=& \frac{1}{2}
\: \delta_{\mu}^{\:\:\:\nu}\: A_{\alpha\beta}\:B^{\alpha\beta}  \ ,\label{ig}
\end{eqnarray}

which is valid for every pair of antisymmetric tensors in a four-dimensional Lorentzian spacetime \cite{MW}, when applied to the case $A_{\mu\alpha} = \xi_{\mu\alpha}$ and $B^{\nu\alpha} = \ast \xi^{\nu\alpha}$ to yield the equivalent condition,

\begin{eqnarray}
\xi_{\alpha\mu}\:\ast \xi^{\mu\nu} &=& 0\ ,\label{i2}
\end{eqnarray}

which is equation (64) in \cite{MW}. It is evident that identity (\ref{i1}) is a special case of (\ref{ig}). The duality rotation given by equation (\ref{dr}) allows us to express the stress-energy tensor in terms of the extremal field,

\begin{equation}
T_{\mu\nu}=\xi_{\mu\lambda}\:\:\xi_{\nu}^{\:\:\:\lambda}
+ \ast \xi_{\mu\lambda}\:\ast \xi_{\nu}^{\:\:\:\lambda}\ .\label{TEMDR}
\end{equation}

With all these elements it becomes trivial to prove that the tetrad (\ref{V1}-\ref{V4}) is orthogonal and diagonalizes locally and covariantly the stress-energy tensor (\ref{TEMDR}). We notice then that we still have to define the vectors $X^{\mu}$ and $Y^{\mu}$. We will now introduce some terms. The tetrad vectors have two essential components. For instance in vector $V_{(1)}^{\alpha}$ there are two main structures. First, the skeleton, in this case $\xi^{\alpha\lambda}\:\xi_{\rho\lambda}$, and second, the gauge vector $X^{\rho}$.  In the case of $V_{(3)}^{\alpha}$, the skeleton will be $\ast \xi^{\alpha\lambda}$, and $Y_{\lambda}$ will be the gauge vector. The gauge vectors as proven in manuscript \cite{A} could be anything that does not make the tetrad vectors trivial. That is, the tetrad (\ref{V1}-\ref{V4}) diagonalizes the stress-energy tensor for any non-trivial gauge vectors $X^{\mu}$ and $Y^{\mu}$. It was therefore proved that we can make different choices for $X^{\mu}$ and $Y^{\mu}$. In geometrodynamics, the Maxwell equations,

\begin{eqnarray}
f^{\mu\nu}_{\:\:\:\:\:;\nu} &=& 0 \label{L1}\nonumber\\
\ast f^{\mu\nu}_{\:\:\:\:\:;\nu} &=& 0 \ , \label{L2}
\end{eqnarray}

show us that two potential vector fields $A_{\nu}$ and $\ast A_{\nu}$ exist \cite{CF},

\begin{eqnarray}
f_{\mu\nu} &=& A_{\nu ;\mu} - A_{\mu ;\nu}\label{ER}\nonumber\\
\ast f_{\mu\nu} &=& \ast A_{\nu ;\mu} - \ast A_{\mu ;\nu} \ .\label{DER}
\end{eqnarray}

The symbol $``;''$ stands for covariant derivative with respect to the metric tensor $g_{\mu\nu}$. The $\ast$ in $\ast A_{\nu}$ is just part of a nomenclature, it does not refer to the Hodge map, meaning that $\ast A_{\nu ;\mu} = (\ast A_{\nu})_{;\mu}$. We can define then, a normalized tetrad with the choice $X^{\mu} = A^{\nu}$ and $Y^{\mu} = \ast A^{\nu}$,

\begin{eqnarray}
U^{\alpha} &=& \xi^{\alpha\lambda}\:\xi_{\rho\lambda}\:A^{\rho} \:
/ \: (\: \sqrt{-Q/2} \: \sqrt{A_{\mu} \ \xi^{\mu\sigma} \
\xi_{\nu\sigma} \ A^{\nu}}\:) \label{U}\\
V^{\alpha} &=& \xi^{\alpha\lambda}\:A_{\lambda} \:
/ \: (\:\sqrt{A_{\mu} \ \xi^{\mu\sigma} \
\xi_{\nu\sigma} \ A^{\nu}}\:) \label{V}\\
Z^{\alpha} &=& \ast \xi^{\alpha\lambda} \: \ast A_{\lambda} \:
/ \: (\:\sqrt{\ast A_{\mu}  \ast \xi^{\mu\sigma}
\ast \xi_{\nu\sigma}  \ast A^{\nu}}\:)
\label{Z}\\
W^{\alpha} &=& \ast \xi^{\alpha\lambda}\: \ast \xi_{\rho\lambda}
\:\ast A^{\rho} \: / \: (\:\sqrt{-Q/2} \: \sqrt{\ast A_{\mu}
\ast \xi^{\mu\sigma} \ast \xi_{\nu\sigma} \ast A^{\nu}}\:) \ .
\label{W}
\end{eqnarray}

The four vectors (\ref{U}-\ref{W}) have the following algebraic properties,

\begin{equation}
-U^{\alpha}\:U_{\alpha}=V^{\alpha}\:V_{\alpha}
=Z^{\alpha}\:Z_{\alpha}=W^{\alpha}\:W_{\alpha}=1 \ .\label{TSPAUX}
\end{equation}

Using the equations (\ref{i1}-\ref{i2}) it is simple to prove that (\ref{U}-\ref{W}) are orthonormal. When we make the transformation,

\begin{eqnarray}
A_{\alpha} \rightarrow A_{\alpha} + \Lambda_{,\alpha}\ , \label{G1}
\end{eqnarray}

$f_{\mu\nu}$ remains invariant, and the transformation,

\begin{eqnarray}
\ast A_{\alpha} \rightarrow \ast A_{\alpha} +
\ast \Lambda_{,\alpha}\ , \label{G2}
\end{eqnarray}

leaves $\ast f_{\mu\nu}$ invariant, as long as the functions $\Lambda$ and $\ast \Lambda$ are scalars. We notice that a local electromagnetic gauge transformation of the ``gauge vectors'' $X^{\alpha}=A^{\alpha}$ and $Y^{\alpha}=\ast A^{\alpha}$ can be just interpreted as a new choice for the gauge vectors $X_{\alpha} = A_{\alpha} + \Lambda_{,\alpha}$ and $Y_{\alpha} = \ast A_{\alpha} + \ast \Lambda_{,\alpha}$. Schouten defined what he called, a two-bladed structure in a spacetime \cite{SCH}. These blades are the planes determined by the pairs ($U^{\alpha}, V^{\alpha}$) and ($Z^{\alpha}, W^{\alpha}$).
It was proved in \cite{A} that the transformation (\ref{G1}) generates a ``rotation'' of the tetrad vectors ($U^{\alpha}, V^{\alpha}$) into ($\tilde{U}^{\alpha}, \tilde{V}^{\alpha}$) provoking that these ``rotated'' vectors ($\tilde{U}^{\alpha}, \tilde{V}^{\alpha}$) remain in the plane or blade one generated by ($U^{\alpha}, V^{\alpha}$). It was also proved in \cite{A} that the transformation (\ref{G2}) generates a ``rotation'' of the tetrad vectors ($Z^{\alpha}, W^{\alpha}$) into ($\tilde{Z}^{\alpha}, \tilde{W}^{\alpha}$) causing that these ``rotated'' vectors ($\tilde{Z}^{\alpha}, \tilde{W}^{\alpha}$) remain in the plane or blade two generated by ($Z^{\alpha}, W^{\alpha}$).  For example, a boost of the two vectors $(U^{\alpha},\:V^{\alpha})$ on blade one, given in (\ref{U}-\ref{V}), by the ``angle'' $\phi$ can be written,

\begin{eqnarray}
U^{\alpha}_{(\phi)}  &=& \cosh(\phi)\: U^{\alpha} +  \sinh(\phi)\: V^{\alpha} \label{UT} \\
V^{\alpha}_{(\phi)} &=& \sinh(\phi)\: U^{\alpha} +  \cosh(\phi)\: V^{\alpha} \label{VT} \ .
\end{eqnarray}

There are also discrete transformations of vectors $(U^{\alpha},\:V^{\alpha})$ on blade one, see reference \cite{A}. The rotation of the two tetrad vectors $(Z^{\alpha},\:W^{\alpha})$ on blade two, given in (\ref{Z}-\ref{W}), by the ``angle'' $\varphi$, can be expressed as,

\begin{eqnarray}
Z^{\alpha}_{(\varphi)}  &=& \cos(\varphi)\: Z^{\alpha} -  \sin(\varphi)\: W^{\alpha} \label{ZT} \\
W^{\alpha}_{(\varphi)}  &=& \sin(\varphi)\: Z^{\alpha} +  \cos(\varphi)\: W^{\alpha} \label{WT} \ .
\end{eqnarray}

A simple algebraic exercise can show us that the equalities $U^{[\alpha}_{(\phi)}\:V^{\beta]}_{(\phi)} = U^{[\alpha}\:V^{\beta]}$ and $Z^{[\alpha}_{(\varphi)}\:W^{\beta]}_{(\varphi)} = Z^{[\alpha}\:W^{\beta]}$ are true. These equalities are telling us that these antisymmetric tetrad objects are gauge invariant. As a reminder we can state that in manuscript \cite{A} it was proven that the group of local electromagnetic gauge transformations is isomorphic to the local group LB1 of boosts plus discrete transformations on blade one, and independently to LB2, the local group of spatial rotations on blade two. Equations (\ref{UT}-\ref{VT}) represent a local electromagnetic gauge transformation of the vectors $(U^{\alpha}, V^{\alpha})$. While equations (\ref{ZT}-\ref{WT}) represent a local electromagnetic gauge transformation of the vectors $(Z^{\alpha}, W^{\alpha})$. Written in terms of these tetrad vectors, the electromagnetic field is,

\begin{equation}
f_{\alpha\beta} = -2\:\sqrt{-Q/2}\:\:\cos\alpha\:\:U_{[\alpha}\:V_{\beta]} +
2\:\sqrt{-Q/2}\:\:\sin\alpha\:\:Z_{[\alpha}\:W_{\beta]}\ .\label{EMF}
\end{equation}

Equation (\ref{EMF}) represents maximum simplification in the electromagnetic field expression. The true degrees of freedom are the local scalars $\sqrt{-Q/2}$ and $\alpha$. Local gauge invariance is manifested explicitly through the possibility of ``rotating'' through a scalar angle $\phi$ on blade one by a local gauge transformation (\ref{UT}-\ref{VT}) the tetrad vectors $U^{\alpha}$ and $V^{\alpha}$, such that
$U_{[\alpha}\:V_{\beta]}$ remains invariant \cite{A}. That is to say, that they remain analogous for discrete transformations on blade one. A similar analysis can be carried out for blade two. A spatial ``rotation'' of the tetrad vectors $Z^{\alpha}$ and $W^{\alpha}$ through an ``angle'' $\varphi$ as in (\ref{ZT}-\ref{WT}), leads $Z_{[\alpha}\:W_{\beta]}$ to remain invariant \cite{A}. This formalism clearly provides a technique to maximally simplify the expression for the electromagnetic field strength. We finally conclude in this brief preview, that through transitivity it can be proven that the boosts plus discrete transformations on plane one are isomorphic to the spatial rotations on plane two. We proceed to apply all this geometrical elements to the Reissner-Nordstr\"{o}m case with the choice $X^{\rho} = A^{\rho}$ and $Y^{\rho} = \ast A^{\rho}$, where the symbol $\ast$ in this particular last case is not the Hodge operator but a particular nomenclature. In the standard spherical coordinates $t, r, \theta, \phi$ the only non-zero components for the potentials will be $A_{t} = - q / r$ and $\ast A_{\phi} = -q\:\cos\theta$. With these potentials we find that the only non-zero components for the electromagnetic tensor $f_{\mu\nu} = A_{\nu ;\mu} - A_{\mu ;\nu}$ and its Hodge dual $\ast f_{\mu\nu} = \ast A_{\nu ;\mu} - \ast A_{\mu ;\nu}$ are $f_{tr} = - q / r^{2}$ and $\ast f_{\theta\phi} = q\:\sin\theta$. The symbol $;$ stands for covariant derivative with respect to the metric tensor $g_{\mu\nu}$, in our case the Reissner-Norstr\"{o}m geometry. It is easy to check that the only non-zero components of the extremal field and its dual are $\xi_{tr} = f_{tr}$ and $\ast\xi_{\theta\phi} = \ast f_{\theta\phi}$. We proceed again to write explicitly the only non-zero components of vectors (\ref{U}-\ref{W}) which are going to be useful when determining the geometric location of all conserved energy-momentum currents,

\begin{eqnarray}
U^{t} &=& - (\sqrt{q^{2}}/q) / \sqrt{1 - {2m \over r} + {q^{2} \over r^{2}}} \label{Ut}\\
V^{r} &=& \sqrt{1 - {2m \over r} + {q^{2} \over r^{2}}} \label{Vr}\\
Z^{\theta} &=& - \sqrt{\cos^{2}\theta} / (r\:\cos\theta)  \label{Ztheta}\\
W^{\phi} &=& -\sqrt{q^{2}}\:\sqrt{\cos^{2}\theta} / (q\:r\:\sin\theta\:\cos\theta) \ .\label{Wphi}
\end{eqnarray}

In this particular coordinate system we would have to be careful because both vectors $V_{(3)}^{\alpha}$ and $V_{(4)}^{\alpha}$ before normalizing would be zero at the coordinate value $\theta = \pi / 2$. As the purpose of this section is not to find suitable coordinate coverings but to show that the conserved currents are vectors inside either blade one or two, we proceed to exhibit these currents and conserved charges taken from reference \cite{JS}. In reference \cite{JS} the energy-momentum currents are defined as $T^{\alpha}_{\:\:\:\beta}\:\xi_{j}^{\beta}$ for $j: 1\cdots4$, where the vectors $\xi_{j}^{\beta}$ are Killing vector fields. These Killing vectors are defined by $\xi_{1} = (1,0,0,0)$, $\xi_{2} = (0,\sin\theta,\cos\phi\:\cot\theta,0)$, $\xi_{3} = (0,\cos\phi,-\sin\phi\:\cot\theta,0)$ and $\xi_{4} = (0,0,0,1)$. The corresponding conserved currents are then,

\begin{eqnarray}
J(\xi_{1}) &=& - {q^{2} \over r^{4}}\: \xi_{1} \label{curr1}\\
J(\xi_{a})  &=&  {q^{2} \over r^{4}}\: \xi_{a}\ .\label{curr234}
\end{eqnarray}

Index $a$ $\in$ ${2,3,4}$. The main objective of this section is to show that the conserved currents are located or belong to either blade one or blade two. We can now check comparing equations (\ref{curr1}-\ref{curr234}) with equations (\ref{Ut}-\ref{Wphi}) that the conserved current $J(\xi_{1})$ belongs or is in the plane determined by the vectors $(U^{\alpha}, V^{\alpha})$, that is blade one, and the other three conserved currents $J(\xi_{a})$ for $a: 2\cdots4$ lie in the orthogonal plane or blade two determined by $(Z^{\alpha}, W^{\alpha})$. The conserved charges are calculated exactly as in reference \cite{JS}. The only one that is non-zero corresponds to the current vector inside blade one and it is given in the case where $m^{2} > q^{2}$ and $r_{+} = m + \sqrt{m^{2} - q^{2}}$ by the value $Q_{1} = 4\:\pi\:q^{2} / r_{+}$. Therefore we have proven our point which states that the four energy-momentum conserved currents belong to either blade one or two. From reference \cite{JS} we can easily see that the Bel currents \cite{JS}$^{-}$\cite{JSSET} are also inside these blades. This finding should not be surprising, see section \ref{appendixI}. A vector inside blade one is invariant under local gauge transformations through the vector $Y^{\rho} = \ast A^{\rho}$, and a vector inside blade two is invariant under local gauge transformations through the vector $X^{\rho} = A^{\rho}$. This is because local Lorentz transformations on blade two do not affect the orthogonal blade one, and viceversa. All geometrical constructions presented in this section are going to help visualize the ideas that in a general setting we are going to present in the next section where we are going to study the dynamics of symmetries under a perturbative scheme.


\section{First order perturbations in geometrodynamics}
\label{fopert}

We introduce first order perturbations to the relevant objects where $\varepsilon$ is an appropriate perturbative parameter,

\begin{eqnarray}
\tilde{g}_{\mu\nu} &=& g_{\mu\nu} + \varepsilon \:h_{\mu\nu} \label{metricfirst}\\
\tilde{\xi}_{\mu\nu}  &=&  \xi_{\mu\nu} + \varepsilon \:\omega_{\mu\nu}\ .\label{extremalfirst}
\end{eqnarray}

The perturbation objects $h_{\mu\nu}$, $\omega_{\mu\nu}$ and the one we are going to introduce next for the electromagnetic tensor are of a physical nature caused by an external agent to the source of preexisting fields. It is worth stressing that they are not the result of a local first order coordinate transformation. We will raise indices with the perturbed metric $\tilde{g}^{\mu\nu} = g^{\mu\nu} - \varepsilon \:h^{\mu\nu}$. We can write the perturbed electromagnetic field through a new local duality transformation as,

\begin{eqnarray}
\tilde{f}_{\mu\nu} &=& \cos\tilde{\alpha}\:\:\tilde{\xi}_{\mu\nu} + \sin\tilde{\alpha}\:\:\ast \tilde{\xi}_{\mu\nu} \ .\label{electromagneticfirst}
\end{eqnarray}

The perturbed local complexion $\tilde{\alpha}$ is not going to be explicitly involved in our analysis.
As done in references \cite{A}$^{,}$\cite{MW} we impose the new condition,

\begin{eqnarray}
\tilde{\xi}_{\mu\nu}\:\:\ast \tilde{\xi}^{\mu\nu} = 0 \ .\label{extremalconditionfirst}
\end{eqnarray}

and through the use of the identity (\ref{ig}), which is valid for every pair of antisymmetric tensors in a four-dimensional Lorentzian spacetime \cite{MW}, we will evidence that when applied to the case $A_{\mu\alpha} = \tilde{\xi}_{\mu\alpha}$ and $B^{\nu\alpha} = \ast \tilde{\xi}^{\nu\alpha}$ it yields the equivalent condition,

\begin{equation}
\tilde{\xi}_{\mu\rho}\:\ast \tilde{\xi}^{\mu\lambda} = 0 \ .\label{scond}
\end{equation}

Even though we are developing a first order perturbative scheme, we avoid writing explicitly the first order approximations, specially in this section, in order to create a general framework that allows to understand the ideas with more clarity. Nonetheless we can display as an explicit example equation (\ref{scond}) that at first order it can be written as,

\begin{equation}
\xi_{\mu\rho}\:\ast \xi^{\mu\lambda} + \varepsilon \:\: (\xi_{\mu\rho}\:\ast \omega^{\mu\lambda} + \omega_{\mu\rho}\:\ast \xi^{\mu\lambda} -
\xi_{\mu\rho}\:\ast \xi_{\sigma\tau}\:h^{\mu\sigma}\:g^{\lambda\tau} - \xi_{\mu\rho}\:\ast \xi_{\sigma\tau}\:h^{\lambda\tau}\:g^{\mu\sigma}) = 0 \ .\label{scondfo}
\end{equation}

The complexion, which is a local scalar, on account of imposing condition (\ref{extremalconditionfirst}) can then be expressed as,

\begin{equation}
\tan(2\tilde{\alpha}) = -  \tilde{f}_{\mu\nu}\:\tilde{f}^{\mu\nu} / \ast \tilde{f}_{\gamma\delta}\:\tilde{f}^{\gamma\delta} \ .\label{complexionfirsttan}
\end{equation}

It will be useful for this matter to write the stress-energy tensor for the perturbed fields (\ref{metricfirst}-\ref{extremalfirst}),

\begin{eqnarray}
\tilde{T}_{\mu\nu}=\tilde{\xi}_{\mu\lambda}\:\:\tilde{\xi}_{\nu}^{\:\:\:\lambda} + \ast \tilde{\xi}_{\mu\lambda}\:\ast \tilde{\xi}_{\nu}^{\:\:\:\lambda} \ . \label{fopstressenergy}
\end{eqnarray}

We next proceed to write the four orthogonal vectors that are going to become an intermediate step in constructing the tetrad that diagonalizes the first order perturbed stress-energy tensor (\ref{fopstressenergy}),

\begin{eqnarray}
\tilde{V}_{(1)}^{\alpha} &=& \tilde{\xi}^{\alpha\lambda}\:\tilde{\xi}_{\rho\lambda}\:X^{\rho}
\label{V1fop}\\
\tilde{V}_{(2)}^{\alpha} &=& \sqrt{-\tilde{Q}/2} \:\: \tilde{\xi}^{\alpha\lambda} \: X_{\lambda}
\label{V2fop}\\
\tilde{V}_{(3)}^{\alpha} &=& \sqrt{-\tilde{Q}/2} \:\: \ast \tilde{\xi}^{\alpha\lambda} \: Y_{\lambda}
\label{V3fop}\\
\tilde{V}_{(4)}^{\alpha} &=& \ast \tilde{\xi}^{\alpha\lambda}\: \ast \tilde{\xi}_{\rho\lambda}
\:Y^{\rho}\ ,\label{V4fop}
\end{eqnarray}

In order to prove the orthogonality of the tetrad (\ref{V1fop}-\ref{V4fop}) it is necessary to use the identity (\ref{ig}) for the case $A_{\mu\alpha} = \tilde{\xi}_{\mu\alpha}$ and $B^{\nu\alpha} = \tilde{\xi}^{\nu\alpha}$, that is,

\begin{eqnarray}
\tilde{\xi}_{\mu\alpha}\:\tilde{\xi}^{\nu\alpha} -
\ast \tilde{\xi}_{\mu\alpha}\: \ast \tilde{\xi}^{\nu\alpha} &=& \frac{1}{2}
\: \delta_{\mu}^{\:\:\:\nu}\:\tilde{Q}\ ,\label{i2t}
\end{eqnarray}

where $\tilde{Q} = \tilde{\xi}_{\mu\nu}\:\tilde{\xi}^{\mu\nu}$ is assumed not to be zero. We also need the condition (\ref{scond}). We are free to choose the vector fields $X^{\alpha}$ and $Y^{\alpha}$, as long as the four vector fields (\ref{V1fop}-\ref{V4fop}) are not trivial. In this section we have essentially proved that we can build for the perturbed fields a replica of our previous formalisms and constructions used for the unperturbed fields. In particular, we are able to write our new local tetrad keeping the same local extremal skeleton structure as in the unperturbed case and define the new local planes of symmetry associated to the perturbed stress-energy tensor. The new local planes of symmetry are going to be tilted with respect to the unperturbed planes.

\section{Dynamical symmetry breaking in geometrodynamics}
\label{dynsymmgeom}

In order to study the notion of symmetry breaking in geometrodynamics we are going to need the results from sections \ref{appendixII} and \ref{appendixIII}. We proceed next to write the first order perturbed covariant derivative of a first order perturbed local contravariant current vector,

\begin{eqnarray}
\tilde{\nabla}_{\mu}\:\tilde{J}^{\lambda} = {\partial \tilde{J}^{\lambda} \over  \partial x^{\mu} } + \Gamma^{\lambda}_{\mu\nu}\:\tilde{J}^{\nu} + \varepsilon \:\tilde{\Gamma}^{\lambda}_{\mu\nu}\:J^{\nu} \ , \label{fopcovdercurr}
\end{eqnarray}

We can rewrite equation (\ref{fopcovdercurr}) as follows,

\begin{eqnarray}
\tilde{\nabla}_{\mu}\:\tilde{J}^{\lambda} = \nabla_{\mu}\:J^{\lambda} + \varepsilon\:{\partial J_{(1)}^{\lambda} \over  \partial x^{\mu} } + \varepsilon \: \Gamma^{\lambda}_{\mu\nu}\:J_{(1)}^{\nu} + \varepsilon \:\tilde{\Gamma}^{\lambda}_{\mu\nu}\:J^{\nu} \ , \label{covdercurrcomp}
\end{eqnarray}

where the first order perturbed local current has been written as $\tilde{J}^{\lambda} = J^{\lambda} + \varepsilon \: J_{(1)}^{\lambda}$. The objective of this section is to show that between an initial constant time hypersurface and an intermediate constant time hypersurface, right when the perturbation starts taking place, the unperturbed local currents are considered to be conserved, that is $\nabla_{\mu}\:J^{\mu} = 0$. This is because the unperturbed energy-momentum current for instance, lies inside the local blade one or blade two, that is to say, the local planes of gauge symmetry. Between the intermediate constant time hypersurface and a final hypersurface the original local current $J^{\mu}$ will be no longer conserved. After the perturbation takes place, the ensuing conservation equation will be $\tilde{\nabla}_{\mu}\:\tilde{J}^{\mu} = 0$ for the perturbed local current. This is because the local planes of symmetry, both blade one and two, will be tilted by the perturbation with respect to the planes on the initial setting. There will be new local planes of symmetry at every point in spacetime. We can see through the new perturbed unnormalized vectors (\ref{V1fop}-\ref{V4fop}) that diagonalize the new perturbed stress-energy tensor (\ref{fopstressenergy}), that the new local planes or blades of symmetry in spacetime after the perturbation took place, will no longer coincide with the old ones. This is the reason why after the perturbations already took place the equation $\nabla_{\mu}\:J^{\mu} = 0$ is no longer valid and according to equation (\ref{covdercurrcomp}) the following result will be correct,

\begin{eqnarray}
\nabla_{\mu}\:J^{\lambda} = - \varepsilon\:{\partial J_{(1)}^{\lambda} \over  \partial x^{\mu} } - \varepsilon \: \Gamma^{\lambda}_{\mu\nu}\:J_{(1)}^{\nu} - \varepsilon \:\tilde{\Gamma}^{\lambda}_{\mu\nu}\:J^{\nu} \ . \label{correctedcurrdiv}
\end{eqnarray}

This is exactly what we might call dynamic symmetry breaking. The old currents $J^{\lambda}$ will be no longer conserved, only the new ones $\tilde{J}^{\lambda}$ will be. Using all the elements of analysis developed so far we will proceed to state the following theorem.

\newtheorem {guesslb1} {Theorem}
\begin{guesslb1}
The local orthogonal planes of symmetry or diagonalization of the stress-energy tensor and associated local groups of tetrad transformations LB1 and LB2 evolve as the continuous perturbation of an external agent takes place. Symmetries are continuously broken and transformed into new symmetries as the local planes of symmetry evolve. \end{guesslb1}
\section{Conclusions}
\label{conclusions}

We have been able to develop the concept of dynamic symmetry breaking for classical electromagnetic fields in a curved four-dimensional Lorentzian spacetime. The analogous notion developed under the realm of Quantum Field Theories for the Standard Model aimed at the creation of mass through a dynamical interacting mechanism \cite{NJL}$^{-}$\cite{SW}. In our work we have produced a dynamical breaking of symmetry through a change in spacetime curvature. The symmetries in the gauge theory of electromagnetic fields are understood through the isomorphisms proved in manuscript \cite{A} as local Lorentz transformations on either blade one or two. These local groups had been named LB1 and LB2. New local tetrad vectors transform inside these blades under the action of these groups. Therefore, symmetry breaking is equivalent to a change in local planes or blades one and two. When an external agent to the preexisting geometry perturbes the original system, the local planes of symmetry are tilted with respect to the original ones. Since there are conserved energy-momentum currents represented by vectors that are locally either inside blade one or blade two as we explicitly proved in the Reissner-Norstr\"{o}m geometry case, and then in a general case in section \ref{appendixII}, symmetry breaking means that these currents will be inside the new local planes of symmetry after the perturbation takes place, which will be the perturbed ones. We assumed that the energy-momentum currents conservation equations are locally invariant either under LB1 or LB2, evident in the Reissner-Norstr\"{o}m geometry case and in a more general aspect for the energy-momentum currents introduced in sections \ref{appendixI} and \ref{appendixII}. The symmetries are going to correspond to new local planes so that the new currents will also be inside the new planes. The old local conservation laws will no longer be held. There will be new ones associated to the new planes of symmetry. The vectors that locally diagonalize the old stress-energy tensor will no longer diagonalize the new perturbed stress-energy tensor. We can specify the old and new tetrad vectors by two features. On one hand what we might call the tetrad vectors skeleton and on the other hand the gauge vectors. As an example of skeleton we can see for instance the $\xi^{\alpha\lambda}\:\xi_{\rho\lambda}$ in the vector $V_{(1)}^{\alpha}$. In the same vector the gauge vector would be $X^{\rho}$. Nonetheless, the local tetrad structure in terms of skeletons, on one hand and in terms of gauge fields on the other, will remain structure invariant after the ensuing perturbation. This constitutes an outstanding property of these new tetrads. We can see this through the two sets of equations (\ref{V1}-\ref{V4}) and (\ref{V1fop}-\ref{V4fop}). Even though the tetrad that diagonalizes the original stress-energy tensor is not the same as the new one for the perturbed stress-energy tensor, the tetrad vectors in both cases are locally structure invariant. In conclusion, in this work we have been able to prove that a change in curvature is associated with a local dynamic symmetry breaking process that we might reinterpret as an evolution of local symmetries into new local symmetries. There is a symmetry evolution, and we evaluate this evolution through the local plane symmetry evolution, or the evolution of blades one and two. In other words, the local evolution of the LB1 and LB2 groups. Of course, we also evaluate the local evolution of energy-momentum currents which accompany the evolution of both local planes of symmetry, to which the ideas of section \ref{appendixII} always apply along the spacetime evolution. It is evident that this whole perturbative scheme can be extended analogously to higher perturbative orders. We quote from \cite{DJG} ``What is missing is a deep understanding of the conceptual framework from which the symmetries and properties of the theory emerge. There are many, many hints that the ultimate formulation of the theory will be extraordinarily rich and deep, but most likely it will look very different from our present, rather primitive, understanding''.

\section{Appendix I}
\label{appendixI}

In this section we will prove that if the locally conserved energy-momentum current $T^{\mu\nu}\:\xi_{\nu}$ satisfies invariance either under the local groups LB1 or LB2, then the vector $\xi^{\mu}$ has to lie either on blade two or blade one respectively. The stress-energy tensor can be written \cite{A},

\begin{equation}
T_{\mu\nu} = (Q/2)\: \left[-U_{\mu}\:U_{\nu} +
V_{\mu}\:V_{\nu} -
Z_{\mu}\:Z_{\nu} - W_{\mu}\:W_{\nu}\right]\ .\label{SET}
\end{equation}

We write the vector field $\xi_{\mu}$ in a general way using the orthonormal tetrad (\ref{U}-\ref{W}),

\begin{eqnarray}
\xi_{\mu} = A\:[\cosh\phi \: U_{\mu} + \sinh\phi \: V_{\mu}] + B\:[\cos\varphi \: Z_{\mu} + \sin\varphi \: W_{\mu}]\ .\label{killing}
\end{eqnarray}

where $A$ and $B$ are local scalars as well as $\phi$ and $\varphi$. Equation (\ref{killing}) represents the superposition of a general vector on blade one and a general vector on blade two. The equation for conservation of the energy-momentum current will be,

\begin{eqnarray}
0 = (T^{\mu\nu}\:\xi_{\nu})_{;\mu} &=& ( T^{\mu\nu}\:\left( \:A\:[ \cosh\phi \: U_{\nu} + \sinh\phi \: V_{\nu} ] \right. \nonumber \\&&
\left. + B\:[ \cos\varphi \: Z_{\nu} + \sin\varphi \: W_{\nu} ]\: \right) \: )_{;\:\mu} \ . \label{conservation}
\end{eqnarray}

Using the orthonormal tetrad vectors (\ref{U}-\ref{W}) and equation (\ref{SET}) we can rewrite (\ref{conservation}) as,

\begin{eqnarray}
0 &=& ( (Q/2)\:\left( \:A\:[ \cosh\phi \: U^{\mu} + \sinh\phi \: V^{\mu} ] \right. \nonumber \\&&
\left. + B\:[ -\cos\varphi \: Z^{\mu} - \sin\varphi \: W^{\mu} ]\: \right) \: )_{;\:\mu} \ . \label{conditions}
\end{eqnarray}

From reference \cite{A} we know that we can produce a full inversion on blade one in expression (\ref{conditions}), $(U^{\mu}, V^{\mu})\rightarrow (-U^{\mu}, -V^{\mu})$.

\begin{eqnarray}
0 &=& ( (Q/2)\:\left( -\:A\:[ \cosh\phi \: U^{\mu} + \sinh\phi \: V^{\mu} ] \right. \nonumber \\&&
\left. + B\:[ -\cos\varphi \: Z^{\mu} - \sin\varphi \: W^{\mu} ]\: \right) \: )_{;\:\mu} \ . \label{conditionsfullinv}
\end{eqnarray}

Adding (\ref{conditions}) and (\ref{conditionsfullinv}) we get,

\begin{eqnarray}
0 &=& ( (Q/2)\:\left( B\:[ -\cos\varphi \: Z^{\mu} - \sin\varphi \: W^{\mu} ]\: \right) \: )_{;\:\mu} \ . \label{conditionsbladetwo}
\end{eqnarray}

Now substracting (\ref{conditions}) and (\ref{conditionsfullinv}) we get,

\begin{eqnarray}
0 &=& ( (Q/2)\:\left( \:A\:[ \cosh\phi \: U^{\mu} + \sinh\phi \: V^{\mu} ] \right) \: )_{;\:\mu} \ . \label{conditionssubstractingone}
\end{eqnarray}

If we now impose current conservation under boosts in expression (\ref{conditionssubstractingone}) we necesarily get $A = 0$.
Therefore, if we impose on conserved currents local gauge invariance under LB1, then the vector $\xi^{\mu}$ must lie on blade two and equation (\ref{conditionsbladetwo}) will be satisfied. Again in expression (\ref{conditions}) we can produce a rotation on blade two $\varphi\rightarrow\varphi+\pi$ and $(\cos\phi, \sin\phi)\rightarrow(-\cos\phi, -\sin\phi)$.

\begin{eqnarray}
0 &=& ( (Q/2)\:\left( \:A\:[ \cosh\phi \: U^{\mu} + \sinh\phi \: V^{\mu} ] \right. \nonumber \\&&
\left. - B\:[ -\cos\varphi \: Z^{\mu} - \sin\varphi \: W^{\mu} ]\: \right) \: )_{;\:\mu} \ . \label{conditionsphipluspi}
\end{eqnarray}

Adding (\ref{conditions}) and (\ref{conditionsphipluspi}) we get,

\begin{eqnarray}
0 &=& ( (Q/2)\:\left( \:A\:[ \cosh\phi \: U^{\mu} + \sinh\phi \: V^{\mu} ] \: \right) \: )_{;\:\mu} \ . \label{conditionsbladeone}
\end{eqnarray}

Now substracting (\ref{conditions}) and (\ref{conditionsphipluspi}) we get,

\begin{eqnarray}
0 &=& ( (Q/2)\:\left( \: B\:[ -\cos\varphi \: Z^{\mu} - \sin\varphi \: W^{\mu} ]\:   \right) \: )_{;\:\mu} \ . \label{conditionssubstractingtwo}
\end{eqnarray}

If we now impose current conservation under general spatial rotations in expression (\ref{conditionssubstractingtwo}) we necesarily get $B = 0$.
Therefore, if we impose local gauge invariance under LB2, then the vector $\xi^{\mu}$ must lie on blade one and equation (\ref{conditionsbladeone}) will be satisfied. Summarizing the results in this section, from (\ref{conditionsbladetwo}) and (\ref{conditionsbladeone}) we conclude that if we impose local gauge invariance either under LB1 or LB2 on the local energy-momentum current conservation equation (\ref{conservation}), the vectors $\xi^{\mu}$ have to lie either on blade two or blade one.

\section{Appendix II}
\label{appendixII}

The Reissner-Nordstr\"{o}m geometry is an exception in the sense that the complexion locally satisfies $\tan(2\:\alpha) = 0.$ The Killing vector fields lie on either plane one or two. Locally, Killing vectors in principle would not have to be vectors lying on either of the two planes if we are talking about geometries other than the Reissner-Nordstr\"{o}m case. Then, the question arises about the existence of locally conserved current vectors lying on either planes in a more general dynamic geometry, for instance where non-null electromagnetic fields are present in a curved four-dimensional spacetime but without the spherical symmetry. That is to say, a spherically symmetric source under the dynamic perturbative action of an external agent as stated at the beginning of this work or specifically in section \ref{fopert}. Electromagnetic and gravitational fields would have to satisfy the Einstein-Maxwell equations even under perturbative interaction. The bottom line is that we are assuming that the perturbed fields $\tilde{g}_{\mu\nu}$, $\tilde{\xi}_{\mu\nu}$ and $\tilde{\alpha}$ will also satisfy the Einstein-Maxwell equations. In this section we will analyze conserved currents for the unperturbed case in general, not necessarily the Reissner-Nordstr\"{o}m geometry case. For the perturbed situation the analysis would be similar by replacing in the Einstein-Maxwell equations for the perturbed fields $\tilde{g}_{\mu\nu}$, $\tilde{\xi}_{\mu\nu}$ and $\tilde{\alpha}$ . One simple way to see that there are always locally conserved currents lying on both planes or blades is the following. If we replace the electromagnetic field in terms of the complexion and the extremal field, see expression (\ref{dr}), inside the Maxwell equations and following reference \cite{MW} equations (62-63), we can see that the extremal field and the complexion must satisfy, in accordance to the Maxwell equations which are a subset of the Einstein-Maxwell equations,

\begin{eqnarray}
\xi^{\mu\nu}_{\:\:\:\:;\nu} &=& - \ast \xi^{\mu\nu}\:\alpha_{\nu} \label{Maxwell1}\\
\ast \xi^{\mu\nu}_{\:\:\:\:;\nu} &=& \xi^{\mu\nu}\:\alpha_{\nu} \ ,\label{Maxwell2}
\end{eqnarray}

where $\alpha_{\nu}$ represents the derivative $\partial \alpha/\partial x^{\nu}$. Therefore we can try and explore the vectors $\xi^{\mu\nu}\:\alpha_{\nu}$ and $\ast\xi^{\mu\nu}\:\alpha_{\nu}$, and see if they are conserved on one hand and if they belong to the planes one and two on the other hand. First we can see that due to the antisymmetry of the extremal field $\xi^{\mu\nu}$ and its dual $\ast \xi^{\mu\nu}$ and to the scalar nature of the complexion $\alpha$ the following equations are satisfied,

\begin{eqnarray}
\xi^{\mu\nu}_{\:\:\:\:;\nu\mu} &=& - (\ast \xi^{\mu\nu}\:\alpha_{\nu})_{;\mu} = 0 \label{LCcurr1}\\
\ast \xi^{\mu\nu}_{\:\:\:\:;\nu\mu} &=& (\xi^{\mu\nu}\:\alpha_{\nu})_{;\mu} = 0 \ .\label{LCcurr2}
\end{eqnarray}

An iterative use of equations (\ref{Maxwell1}-\ref{Maxwell2}) leads to equations (\ref{LCcurr1}-\ref{LCcurr2}). If the geometry is such that the complexion gradient $\alpha_{\nu}$ is not trivial, then we have two conserved local vector fields. Next we would like to see for instance if the vector $\xi^{\mu\nu}\:\alpha_{\nu}$ lies on plane one. Using property (\ref{i2}) and the normalized tetrad (\ref{U}-\ref{W}) it is evident to see that it lies on the plane generated by the vectors (\ref{U}-\ref{V}), that is blade one. A similar line of thinking for the vector $\ast\xi^{\mu\nu}\:\alpha_{\nu}$ on blade two. We can summarize these results in the following table,

\begin{eqnarray}
U_{\alpha}\:\xi^{\alpha\beta} &=& \sqrt{-Q/2}\:\:V^{\beta}\label{} \label{projU} \\
V_{\alpha}\:\xi^{\alpha\beta} &=& \sqrt{-Q/2}\:\:U^{\beta} \label{projV} \\
Z_{\alpha}\:\ast \xi^{\alpha\beta} &=& \sqrt{-Q/2}\:\:W^{\beta} \label{projZ} \\
W_{\alpha}\:\ast \xi^{\alpha\beta} &=& -\sqrt{-Q/2}\:\:Z^{\beta}\ . \label{projW}
\end{eqnarray}

Due to property (\ref{i2}) all other contractions are null. We can also observe that we can write the conserved currents as $T^{\mu\nu}\:\xi_{\nu}$. Using the property $T_{\mu\nu}\:T^{\gamma\nu} = (Q/2)^{2}\:\delta_{\mu}^{\:\:\gamma}$ we can find $\xi^{\mu} = \xi^{\mu\nu}\:\alpha_{\nu} / (Q/2)$ on blade one or $\xi^{\mu} = \ast \xi^{\mu\nu}\:\alpha_{\nu} / (Q/2)$ on blade two. The vector $\xi^{\mu}$ does not have to be necessarily a Killing vector field, nonetheless the energy-momentum current $T^{\mu\nu}\:\xi_{\nu}$ is going to be conserved. Therefore, we proved our point. As long as the gradient of the complexion is not trivial or its contraction with the extremal field and its dual are not trivial, we always have a local conserved current on blade one and another one on blade two. By always we mean during the dynamical evolution. We would like to briefly remind that when a perturbative treatment is implemented, the vacuum Einstein-Maxwell equations can be written as,

\begin{equation}
\widehat{E}(g_{\alpha\beta})=
\widehat{E}(g_{\alpha\beta}^{(0)}+\varepsilon\,g_{\alpha\beta}^{(1)}+
\frac{1}{2}\varepsilon^{2}\,g_{\alpha\beta}^{(2)}+\cdots, \xi_{\alpha\beta}^{(0)}+\varepsilon\,\xi_{\alpha\beta}^{(1)}+
\frac{1}{2}\varepsilon^{2}\,\xi_{\alpha\beta}^{(2)}+\cdots, \alpha^{(0)}+\varepsilon\,\alpha^{(1)}+
\frac{1}{2}\varepsilon^{2}\,\alpha^{(2)}+\cdots)=0
\label{E}
\end{equation}

where $\widehat{E}$ represents the set of nonlinear differential
operators that generates the Einstein-Maxwell equations.  The terms in
(\ref{E}) that are zero order in $\varepsilon$ will be satisfied
automatically because $g_{\alpha\beta}^{(0)}$, the background metric, $\xi_{\alpha\beta}^{(0)}$ the background extremal field and $\alpha^{(0)}$
are a solution to the vacuum Einstein-Maxwell equations.  To find the equations
satisfied by the first order perturbation, we expand (\ref{E}) in
powers of $\varepsilon$, and write the set of first order equations as
\begin{equation}
\widehat{O}(g_{\alpha\beta}^{(1)}, \xi_{\alpha\beta}^{(1)}, \alpha^{(1)})=0\ .
\label{FO}
\end{equation}
Since the perturbations $g_{\alpha\beta}^{(1)}$, $\xi_{\alpha\beta}^{(1)}$ and $\alpha^{(1)}$ can only appear
linearly, $\widehat{O}$ represents a set of linear differential operators. In the set of equations (\ref{E}) we include all the Eistein-Maxwell equations. It is clear that all the analysis done through equations (\ref{Maxwell1}-\ref{Maxwell2}) and (\ref{LCcurr1}-\ref{LCcurr2}) can be reproduced analogously for the perturbed fields $\tilde{g}_{\mu\nu}$, $\tilde{\xi}_{\mu\nu}$ and  $\tilde{\alpha}$. Perturbed fields will arise during the dynamical interaction process.

\section{Appendix III}
\label{appendixIII}

In order to compare local current conservation laws we are going to need the first order perturbed covariant derivative of a vector. In this section we will display the main steps to obtain these calculations. We can start with the standard expression for the covariant derivative of a vector,

\begin{eqnarray}
V^{\lambda}_{\:\:\:\:;\mu} = {\partial V^{\lambda} \over  \partial x^{\mu} } + \Gamma^{\lambda}_{\mu\nu}\:V^{\nu}   \ , \label{covder}
\end{eqnarray}

where the expression for the affine connection is the usual,

\begin{eqnarray}
\Gamma^{\lambda}_{\mu\nu} = {1 \over 2}\:g^{\lambda\sigma}\:\left( {\partial g_{\mu\sigma} \over \partial x^{\nu}} +
{\partial g_{\nu\sigma} \over \partial x^{\mu}} - {\partial g_{\mu\nu} \over \partial x^{\sigma}} \right)\ . \label{affconn}
\end{eqnarray}

Following the literature in perturbative schemes, see \cite{WE}$^{-}$\cite{GP} and references therein as examples, we can write the first order perturbed affine connection as,

\begin{eqnarray}
\tilde{\Gamma}^{\lambda}_{\mu\nu} = {1 \over 2}\:g^{\lambda\sigma}\:\left( h_{\mu\sigma\:;\nu} + h_{\nu\sigma\:;\mu} - h_{\mu\nu\:;\sigma} \right)\ , \label{fopaffconn}
\end{eqnarray}

where the covariant derivatives in (\ref{fopaffconn}) are calculated with the unperturbed (\ref{affconn}) affine connection. We proceed then to write to first order the perturbed covariant derivative of a perturbed contravariant vector,

\begin{eqnarray}
\tilde{\nabla}_{\mu}\:\tilde{V}^{\lambda} = {\partial \tilde{V}^{\lambda} \over  \partial x^{\mu} } + \Gamma^{\lambda}_{\mu\nu}\:\tilde{V}^{\nu} + \varepsilon \:\tilde{\Gamma}^{\lambda}_{\mu\nu}\:V^{\nu} \ , \label{fopcovder}
\end{eqnarray}

where we have used now the operator $\nabla$ to indicate covariant derivative for notational convenience since we can write a tilde above it. The perturbed vector can be written $\tilde{V}^{\lambda} = V^{\lambda} + \varepsilon \: \psi^{\lambda}$, where $\psi^{\lambda}$ is a local vector field. When we think of $V^{\lambda}$ in a concrete example in this manuscript, we will be thinking of the local currents $J^{\lambda}$. It is important to stress that we are studying genuine physical perturbations to the gravitational and electromagnetic fields by external agents to the preexisting source. We are not introducing first order coordinate transformations of the kind $\tilde{x}^{\alpha} = x^{\alpha} + \:\varepsilon \:\:\: \zeta^{\alpha}$, where the local vector field $\zeta^{\alpha}(x^{\sigma})$ is associated to a first order infinitesimal local coordinate transformation scheme \cite{WE}.


\end{document}